\documentclass[12pt,preprint]{emulateapj}

\usepackage{longtable}

\shorttitle{Broad-Line AGN Host Galaxies}
\shortauthors{Trump et al.}

\begin{document}


\title{A Census of Broad-Line Active Galactic Nuclei in Nearby
  Galaxies: Coeval Star Formation and Rapid Black Hole Growth}

\author{
  Jonathan R. Trump,\altaffilmark{1}
  Alexander D. Hsu,\altaffilmark{2}
  Jerome J. Fang,\altaffilmark{1}
  S. M. Faber,\altaffilmark{1}
  David C. Koo,\altaffilmark{1}
  and Dale D. Kocevski\altaffilmark{1}
}

\altaffiltext{1}{
  University of California Observatories/Lick Observatory and
  Department of Astronomy and Astrophysics, University of California,
  Santa Cruz, CA 95064 USA
\label{UCO/Lick}}

\altaffiltext{2}{
  The Harker School, 500 Saratoga Avenue, San Jose, CA 95129 USA
\label{Arizona}}

\def\etal{et al.}
\newcommand{\Ha}{\hbox{{\rm H}$\alpha$}}
\newcommand{\Hb}{\hbox{{\rm H}$\beta$}}
\newcommand{\MgII}{\hbox{{\rm Mg}\kern 0.1em{\sc ii}}}
\newcommand{\CIII}{\hbox{{\rm C}\kern 0.1em{\sc iii}]}}
\newcommand{\CIV}{\hbox{{\rm C}\kern 0.1em{\sc iv}}}
\newcommand{\OII}{\hbox{[{\rm O}\kern 0.1em{\sc ii}]}}
\newcommand{\OIII}{\hbox{[{\rm O}\kern 0.1em{\sc iii}]}}
\newcommand{\OIV}{\hbox{[{\rm O}\kern 0.1em{\sc iv}]}}
\newcommand{\NII}{\hbox{[{\rm N}\kern 0.1em{\sc ii}]}}

\begin{abstract}
  We present the first quantified, statistical map of broad-line
  active galactic nucleus (AGN) frequency with host galaxy color and
  stellar mass in nearby ($0.01<z<0.11$) galaxies.  Aperture
  photometry and $z$-band concentration measurements from the Sloan
  Digital Sky Survey (SDSS) are used to disentangle AGN and galaxy
  emission, resulting in estimates of uncontaminated galaxy rest-frame
  color, luminosity, and stellar mass.  Broad-line AGNs are
  distributed throughout the blue cloud and green valley at a given
  stellar mass, and are much rarer in quiescent (red sequence)
  galaxies.  This is in contrast to the published host galaxy
  properties of weaker narrow-line AGNs, indicating that broad-line
  AGNs occur during a different phase in galaxy evolution.  More
  luminous broad-line AGNs have bluer host galaxies, even at fixed
  mass, suggesting that the same processes that fuel nuclear activity
  also efficiently form stars.  The data favor processes that
  simultaneously fuel both star formation activity and rapid
  supermassive black hole accretion.  If AGNs cause feedback on their
  host galaxies in the nearby universe, the evidence of galaxy-wide
  quenching must be delayed until after the broad-line AGN phase.
\end{abstract}


\section{Introduction}

The well-studied correlations between supermassive black hole (SMBH)
mass and properties of the host galaxy bulge \citep[e.g.,][]{mag98,
  geb00, fer00, mar03} indicate that SMBH growth may be intimately
tied to galaxy evolution.  However the causal physics behind the
SMBH-galaxy link remains mysterious.  Theoretical simulations suggest
that rapidly accreting SMBHs in the active galactic nucleus (AGN)
phase correspond to periods of recent massive star formation in their
host galaxies \citep[e.g.,][]{hop06, hop08}.  Powerful AGNs may also
cause ``feedback'' on their host galaxies, shutting down star
formation by blowing out the star-forming gas either via radiative
winds \citep{sil98,fab02,dim05} or radio jets \citep{cro06}.
Observational evidence for star formation or feedback coevolving with
black hole growth can be found in the star formation histories of AGN
host galaxies \citep[e.g.,][]{hec06}.

Of particular interest are the hosts of the most luminous AGNs, which
are readily identified by broad emission lines in their optical
spectra \citep[e.g.,][]{van01}.  These broad-line AGNs rapidly accrete
material at rates of 1\% to 100\% of the Eddington limit
\citep{kol06,tru09}, and so require plentiful gas in their hosts, with
the possibility of accompanying star formation activity.  There is
also evidence that broad-line AGNs universally have high velocity
optical and X-ray outflows \citep{gan08,win10}, indicative of powerful
winds and the potential for effective feedback in shutting down star
formation.  Studying the mass and rest-frame color of the host galaxy
can reveal its recent star formation history: at a given stellar mass,
galaxies that are very blue in rest-frame $u-z$ have recently
experienced a great deal of star formation, while red galaxies are
quiescent and dominated by old stars.  However the brightness of many
broad-line AGNs complicates observations of their host galaxies, since
the AGN often outshines the galaxy's starlight.

Most authors simply avoid the problem of AGN contamination by studying
host-dominated AGN.  ``Host-dominated'' means that the accreting black
hole is either obscured or weakly accreting (Eddington ratios of
$<$1\%), and the photometry is dominated by the host galaxy.  Studies
of such host-dominated AGNs suggest a preference for massive
($\log(M_*/M_\odot)>10.5$) host galaxies \citep{kau03b,hag10},
although \citet{aird12} suggest this is a relic of selection effects
and AGNs are instead equally likely to be in hosts of any stellar
mass.  Several observations suggest that host-dominated AGNs are most
often found in ``green valley'' galaxies, so-called because they have
colors intermediate between the more densely populated star-forming
blue cloud and the passive red sequence of galaxies
\citep{nan07,sal07,geo08,sil08,gab09,schaw09,hic09,koc09}.  Other
studies, however, argue that active galaxies have the same color
distribution as inactive galaxies of similar mass and the apparent
green valley peak for AGN hosts is caused only because AGNs prefer
massive galaxies \citep{sil09,xue10}.  \citet{car10} additionally
argue that the apparent green valley hosts of AGNs are just
dust-reddened star-forming (and intrinsically blue) galaxies, and
dust-corrected AGN hosts have the same color distribution as inactive
galaxies.

Besides the difficulties in the differing interpretations, the above
studies include only obscured or weakly accreting AGNs with very
different fueling and outflow properties from luminous broad-line AGNs
\citep{ho08,tru11}.  By definition, host-dominated AGNs do not
dominate the energetics of their hosts, and probably have minimal
influence on the current star formation in their galaxies.  Studying
the impact of AGNs on galaxy evolution requires observing galaxies
during the most rapid period of SMBH growth.

Studies of broad-line AGN host galaxies have generally used structural
decomposition of high spatial resolution {\it Hubble Space Telescope}
({\it HST}) images \citep[e.g.,][]{peng02}.  By modeling the luminous
point-source AGN and subtracting it from the extended host galaxy
light, the intrinsic galaxy properties can be disentangled from the
contaminating AGN.  The first high-resolution studies of quasar hosts
suggested that these luminous AGNs prefer massive and luminous hosts
\citep{bah97}.  The host galaxies of luminous AGNs were also found to
have younger stellar populations than inactive galaxies of the same
mass, with colors spanning the blue cloud and green valley
\citep{jah04a,jah04b}.  However the need for high-resolution {\it HST}
imaging limited these studies to small numbers ($\sim$20) of AGNs.
Recent Herschel far-infrared studies, also limited to small samples,
similarly suggest that more luminous AGNs have higher star formation
rates than their inactive counterparts (\citealp{san12,rov12};
\citealp[but see also][]{mul12}).

In this work we expand host galaxy studies of rapidly accreting AGNs
using 820 broad-line AGNs at $0.01<z<0.11$ from the Sloan Digital Sky
Survey.  The large number of AGNs, with a set of matched inactive
galaxies, allows for the first statistical map of broad-line AGN
frequency across the nearby galaxy color-mass diagram.  The selection
and properties of the data are described in Section 2.  Our AGN/host
decomposition method is described in Section 3, which introduces a
novel aperture photometry method for separating point-source AGNs and
extended inactive galaxies.  Section 4 reveals that broad-line AGNs
have a strong preference for star-forming galaxies, and demonstrates
that this preference is not a function of selection effects.  We
discuss what these results mean for the relationship between nuclear
activity and star formation, and the efficacy of AGN feedback, in
Section 5.

Throughout this work we assume a cosmology with $h=0.70$,
$\Omega_M=0.3$, $\Omega_{\Lambda}=0.7$.

\section{Observational Data}

We select samples of broad-line AGNs and inactive galaxies from the
Sloan Digital Sky Survey \citep[SDSS,][]{york00}.  Each type is
selected using the spectroscopic classification provided by the SDSS
DR7: broad-line AGNs are identified by {\it SpecClass}$=3$, while
galaxies without broad emission lines have {\it SpecClass}$=2$.  We
visually inspected the spectra of all broad emission line AGNs to
ensure they are correctly classified (removing the misclassified
$\sim$1\% of objects).  Note that we refer to all galaxies without
broad lines as inactive, while some of these ``inactive'' galaxies may
actually have emission line ratios that suggest weak or obscured AGNs
\citep{bpt81,kew06}.

Given the parent sample of broad-line AGNs and inactive
(non-broad-line) galaxies in the SDSS, we make the following cuts:

\begin{enumerate}
  \item Face-on systems only, with $b/a>0.5$ (where $b/a$ is the ratio
    between the minor and major axes of the $r$-band image).  This
    constraint is designed to eliminate dusty systems and removes a
    significant number (36\%) of inactive galaxies but only 15\% of
    the initial broad-line AGNs.\footnote{The high $b/a$ preference
      for broad-line AGNs could be a selection bias, such that a point
      source is biasing the axis ratio measurement.  However several
      studies also suggest a genuine preference for broad-line AGNs to
      lie in spheroid-dominated \citep{bah97} or face-on \citep{rig06}
      galaxies.}

  \item $0.01<z<0.11$.  The sample is further sub-divided into
    $\Delta{z}=0.01$ bins (i.e., $0.01<z<0.02$, $0.02<z<0.03$, etc.)
    for the AGN light correction, as described in Section 3.

  \item $r<17.77$.  This is the spectroscopy limit for inactive
    galaxies in the SDSS.  Although the SDSS spectroscopy includes
    broad-line AGNs to fainter magnitudes ($i<19.1$ or $i<20.2$), we
    require $r<17.77$ for both samples to ensure a complete control
    sample.
\end{enumerate}

The SDSS spectroscopy is $>$95\% complete to both galaxies and quasars
in these redshift and magnitude ranges \citep{str02,ric02}.  These
initial selection criteria result in a total of 192,946 inactive
galaxies and 972 broad-line AGNs.  Beyond the redshift and magnitude
limits, the sample is divided into luminosity-limited ``faint'' and
``luminous'' samples, each complete to a given $r$-band absolute
magnitude.

\begin{itemize}
  \item Luminous sample: This is limited to all $0.01<z<0.11$ sources
    with $M_r'<-20.8$, which corresponds to the $r<17.77$ spectroscopy
    limit at $z=0.11$.  The luminous sample probes the full redshift
    range of the sample, and includes 769 broad-line AGNs and 110,670
    inactive galaxies.

  \item Faint sample: This is limited to all $0.01<z<0.05$ sources
    with $M_r'<-19$, which similarly corresponds to the $r<17.77$
    limit at $z=0.05$.  The faint sample is designed to probe a large
    range of luminosities and stellar masses, with 134 broad-line AGNs
    and 33,361 inactive galaxies.
\end{itemize}

Note that galaxies in the faint sample with $M_r'<-20.8$ are also in
the luminous sample, and there are a total of 820 unique broad-line
AGNs in the two samples.

\subsection{Photometry}

The SDSS provides magnitudes in five $ugriz$ filters for all
broad-line AGNs and inactive galaxies.  We use both the total
magnitude $m$ integrated across the entire galaxy\footnote{Total
  magnitude $m$ is the SDSS ``model magnitude.''  These magnitudes are
  computed over the same aperture in all five filters, with the
  aperture size computed from the model fit to the $r$ band.}, and the
inner aperture magnitude $m_{in}$ measured within a 3\arcsec~diameter
of the galaxy center\footnote{The inner aperture magnitude is not to
  be confused with the SDSS spectroscopic fiber magnitude: $m_{in}$ is
  measured from the same photometric image as $m$.}.  An outer
aperture magnitude $m_{out}$ is calculated from the light outside the
3\arcsec~diameter, given by
\begin{equation}
  m_{out} = -2.5\log( 10^{-0.4m}-10^{-0.4m_{in}} ).
\end{equation}
Note that $m_{in}$ is calculated after convolving the image to
2\arcsec, which ensures uniform seeing for all objects.  While
convolving to lower resolution slightly worsens AGN contamination, it
actually helps our AGN/galaxy decomposition by ensuring that all
active galaxies have the same 2\arcsec~resolution as the point-source
stars and inactive galaxies used to calibrate the method.

\begin{figure}[t]
\scalebox{1.2}
{\plotone{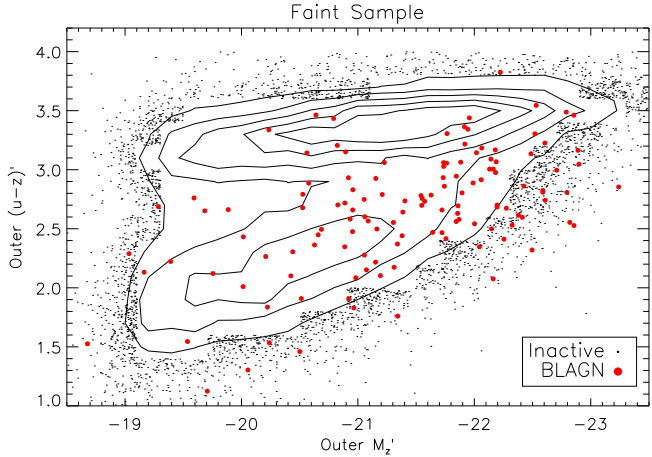}}
\figcaption{Outer $(u-z)'$ color with outer $M_z'$ absolute magnitude,
  each K-corrected to $z=0.05$, for both inactive galaxies (contours
  and points) and broad-line AGNs (filled red circles) in the
  faint sample.  Outer magnitudes represent the light outside a
  3\arcsec~diameter aperture.  Broad-line AGNs lie all over the
  color-magnitude diagram but tend to appear brighter and bluer than
  inactive galaxies because the AGN light contaminates even the outer
  magnitude measurements.  Section 3 outlines the derivation and
  application of a correction that recovers the uncontaminated host
  galaxy light for broad-line AGNs.
\label{fig:colormag_faint}}
\end{figure}

\begin{figure}[t]
\scalebox{1.2}
{\plotone{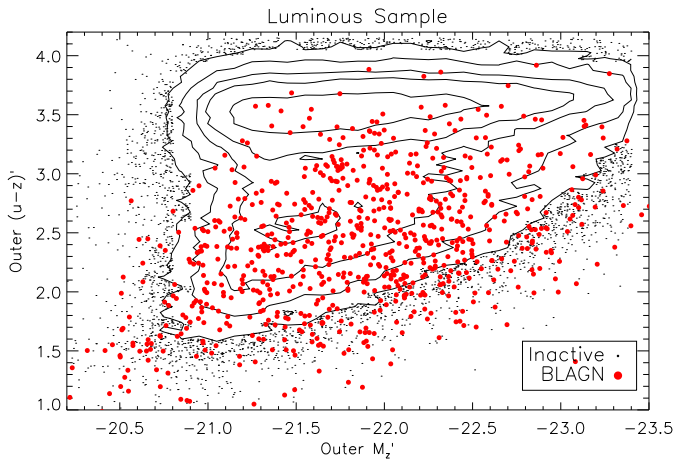}}
\figcaption{K-corrected outer $(u-z)'$ color with outer $M_z'$
  absolute magnitude for the luminous sample, where outer magnitudes
  are defined as the light outside a 3\arcsec~diameter aperture.  As
  in Figure \ref{fig:colormag_faint}, light from the AGN causes their
  hosts to appear brighter and bluer than inactive galaxies.  The
  contaminating blue AGN light is particularly evident in the many
  AGNs with total $M_r'<-20.8$ but outer $M_z'>-20.8$ (in contrast to
  the typically redder inactive galaxy population).
\label{fig:colormag_lumin}}
\end{figure}

We K-correct the observed photometry in both samples to the $z=0.05$
frame using the public {\tt kcorrect} IDL software \citep{kcorrect}.
The prime ($'$) notation is used to denote colors and magnitudes
K-corrected to $z=0.05$.  Figure \ref{fig:colormag_faint} shows the
K-corrected outer $(u-z)'$ color with the outer $M_z'$ absolute
magnitude for inactive galaxies and broad-line AGNs in the faint
sample, and Figure \ref{fig:colormag_lumin} similarly shows the
luminous sample.  Naively one might assume that the outer magnitudes
do not contain any light from the AGN point source.  However both
figures show that many broad-line AGNs have brighter and bluer outer
magnitudes than the inactive galaxy population.  Light from the SDSS
point spread function (PSF) extends beyond the inner 3\arcsec
aperture, and a more sophisticated process is necessary to recover the
uncontaminated galaxy properties.  This technique is derived and
applied in Section 3.

\subsection{Stellar Masses}

Galaxy stellar masses come from the MPA-JHU value-added catalog,
derived according to \citet{kau03a}.  Derived masses for the inactive
galaxies have $1\sigma$ errors of $\sim$0.05~dex, with no error
dependence on mass or magnitude.  However most of the broad-line AGN
hosts are absent from this catalog, and those included in the catalog
probably have inaccurate stellar masses due to the AGN contamination.
To estimate masses for AGN hosts, we first calculate the mass-to-light
ratios of inactive galaxies as a function of color and luminosity.

\begin{figure}[t]
\scalebox{1.2}
{\plotone{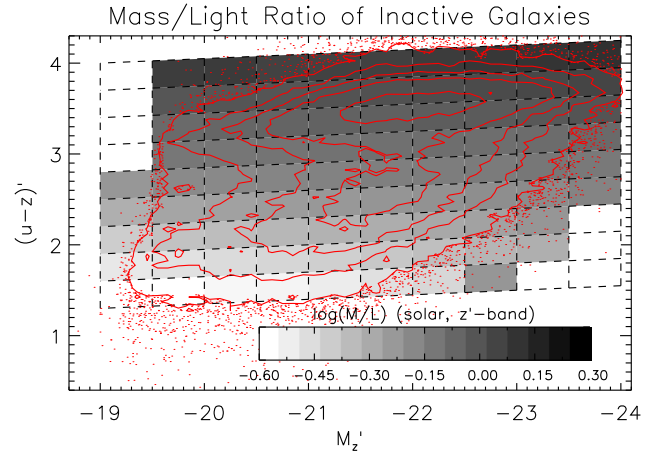}}
\figcaption{Median mass-to-light ratio, in the $z'$-band, for all
  inactive galaxies with $r<17.77$ at $0.01<z<0.11$.  Mass-to-light
  ratio is computed in bins of color $(u-z)'$ and luminosity ($M_z'$),
  with no data shown for bins containing fewer than 5 galaxies.
  Mass-to-light ratio is a strong function of color and a weak
  function of luminosity.  This figure is used to estimate masses for
  broad-line AGN hosts, using their corrected (AGN-subtracted) galaxy
  colors and luminosities to determine the appropriate mass-to-light
  ratio.
\label{fig:masslight}}
\end{figure}

Figure \ref{fig:masslight} shows the median $z'$-band mass-to-light
ratio in bins across the color-magnitude diagram.  We use this figure
to estimate masses for broad-line AGN hosts, applying the
mass-to-light ratio from the bin corresponding to their corrected
(AGN-subtracted) host galaxy color and absolute magnitude.

\section{Disentangling AGN and Galaxy Light}

Broad-line AGNs dominate the light in the inner \mbox{($<$3$\arcsec$)}
aperture but also contribute light in the outer \mbox{($>$3$\arcsec$)}
aperture (as shown by Figures \ref{fig:colormag_faint} and
\ref{fig:colormag_lumin}).  We seek to remove the AGN emission to
obtain uncontaminated measurements of host galaxy light.  It turns out
that inactive galaxies, uncontaminated by AGN light, have fairly tight
relationships between $z$-band concentration and inner and outer
magnitudes in each filter: we exploit these relationships to predict
the galaxy-only magnitudes of AGN hosts.  Likewise the relationship
between inner and outer magnitudes of stars can predict the magnitudes
of the AGN-only point-source component.

\subsection{Galaxy Inner and Outer Aperture Magnitudes}

\begin{figure}[t]
\scalebox{1.18}
{\plotone{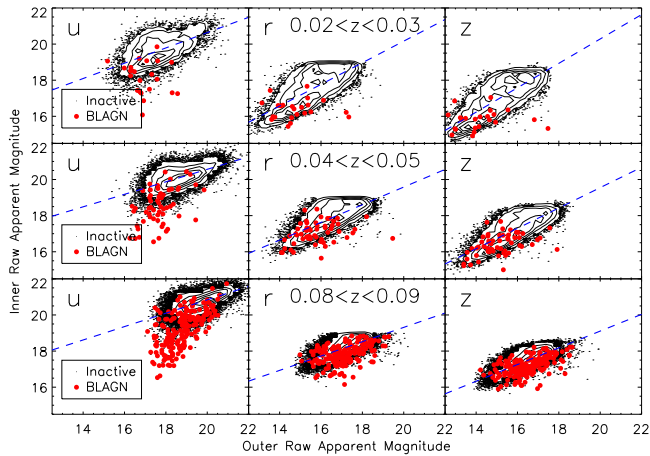}}
\figcaption{Inner vs. outer magnitudes for inactive galaxies (black
  contours and points) and broad-line AGNs (filled red circles) in
  three different filters and redshift ranges.  The dashed blue line
  shows the best-fit line to the inactive galaxy population.  Galaxies
  are typically fainter in $u$ and $r$ than $z$, and more distant
  galaxies appear fainter in their outer magnitudes (as their smaller
  apparent sizes cause more light to fall in the inner aperture).
  Broad-line AGN hosts typically have brighter inner magnitudes,
  particularly in blue light.  The brightest AGNs additionally have
  slightly brighter outer magnitudes than inactive galaxies, as some
  AGN light ``leaks'' beyond the inner aperture and contaminates the
  outer aperture light (especially in the $u$ band).  The offset of a
  galaxy from the best-fit line is the residual ${\Delta}m$ in
  Equation \ref{eq:deltamag}.
\label{fig:innerouter}}
\end{figure}

We begin by comparing the inner \mbox{($<$3$\arcsec$)} and outer
\mbox{($>$3$\arcsec$)} aperture magnitudes of inactive galaxies and
broad-line AGNs.  Figure \ref{fig:innerouter} shows inner and outer
magnitudes in three different filters ($u$, $r$, $z$) and redshift
ranges ($0.02<z<0.03$, $0.04<z<0.05$, $0.08<z<0.09$).  Light from
broad-line AGNs has the strongest contaminating effect in blue light
and inner magnitudes, but also affects outer magnitudes.  For each of
the five filters and in ten bins of redshift ($0.01<z<0.11$ in
${\Delta}z=0.01$ intervals), we find the best-fit line describing
outer ($m_{out,GAL}$) and inner magnitude ($m_{in,GAL}$) for inactive
galaxies (shown by the dashed lines for the filters and redshift
ranges in Figure \ref{fig:innerouter}).  The offset above this line is
defined by:
\begin{equation}
  {\Delta}m = A + B m_{out,GAL} - m_{in,GAL} \label{eq:deltamag}.
\end{equation}

The best-fit line is given by ${\Delta}m=0$, and broad-line AGNs
typically have ${\Delta}m>0$.  However even inactive galaxies have a
large scatter in ${\Delta}m$, presumably because there is significant
structural variation in galaxies of a given outer apparent magnitude.
Adding a structural measurement could better describe the typical
relationship for inner and outer magnitudes in inactive galaxies and
explain their scatter about ${\Delta}m=0$.  In particular we use
$z$-band concentration, defined as the ratio between the radius
containing 90\% of the $z$-band light and the radius containing 50\%
of the light, $C_z=R_{90,z}/R_{50,z}$.

\begin{figure}[t]
\scalebox{1.18}
{\plotone{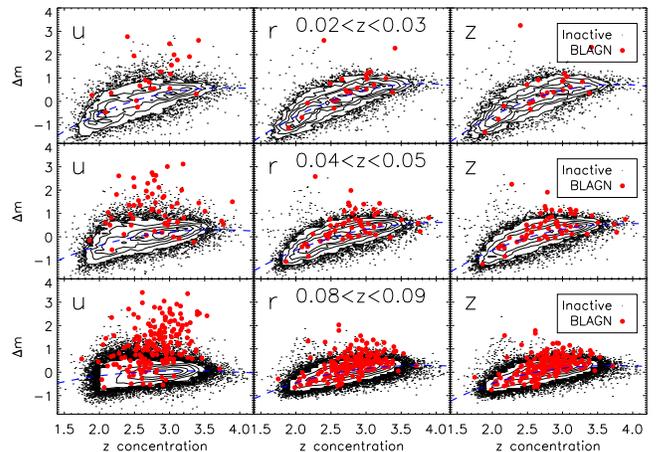}}
\figcaption{Offset from the best-fit line describing inner and outer
  magnitudes in Figure \ref{fig:innerouter} (${\Delta}m$, given by
  Equation \ref{eq:deltamag}) versus $z$-band concentration ($C_z$)
  for inactive galaxies (black contours and points) and broad-line
  AGNs (filled red circles).  The panels show the same three filters
  and redshift ranges as in Figure \ref{fig:innerouter}.  The dashed
  blue line shows the best-fit cubic line to the inactive galaxy
  population.  Broad-line AGNs have the same range in $C_z$ as
  inactive galaxies, but tend to have significantly higher ${\Delta}m$
  (especially in blue light).  The offset from the best-fit line is
  the residual ${\delta}m$ in Equation \ref{eq:cdeltamag}.
\label{fig:deltamag}}
\end{figure}

Figure \ref{fig:deltamag} shows ${\Delta}m$ from Equation
\ref{eq:deltamag} versus $z$-band concentration $C_z$.  Galaxies with
${\Delta}m<0$ (from fainter inner magnitudes than the average given
their outer magnitude) have low concentration, while galaxies with
${\Delta}m>0$ (and brighter inner magnitudes) are more concentrated.
The inactive population shows a tighter relation after using $C_z$ to
describe the structural variation, and broad-line AGNs scatter to
brighter inner magnitudes and higher ${\Delta}m$.

A cubic line is fit to the inactive galaxies in Figure
\ref{fig:deltamag} (shown by the dashed line), with the offset from
this line given by:
\begin{equation}
  {\delta}m = {\Delta}m - C - D C_z - E C_z^2 - F C_z^3.
    \label{eq:cdeltamag}
\end{equation}

\begin{figure}[t]
\scalebox{1.18}
{\plotone{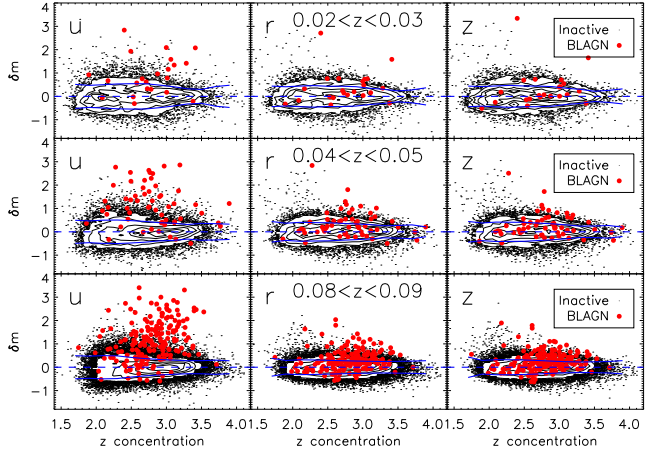}}
\figcaption{Offset from the best-fit cubic describing galaxy structure
  in Figure \ref{fig:deltamag} (${\delta}m$, given by Equation
  \ref{eq:cdeltamag}) versus $z$-band concentration ($C_z$) in the
  same filters and redshift ranges as in Figures \ref{fig:innerouter}
  and \ref{fig:deltamag}.  Inactive galaxies are shown by black
  contours and points and broad-line AGN hosts are given by filled red
  circles.  The dashed blue line shows ${\delta}m=0$, and the solid
  blue lines show the scatter of the inactive galaxies about
  ${\delta}m=0$.
\label{fig:cdeltamag}}
\end{figure}

Figure \ref{fig:cdeltamag} shows ${\delta}m$ with $C_z$ for inactive
galaxies and AGNs.  Broad-line AGN hosts tend to have ${\delta}m>0$,
as the AGN light causes the inner aperture magnitude to be brighter
than expected given the galaxy's $z$ concentration.  Inactive galaxies
have ${\delta}m \sim 0$ with small scatter: the standard deviation is
typically only $\sigma_{{\delta}m}=0.3$~mag, with slightly larger
$\sigma_{{\delta}m}=0.5$~mag scatter for low-concentration ($C_z<2.5$)
galaxies in the $u$ band.  We use ${\delta}m$ as an estimate of the
brightness excess in the inner aperture due to AGN light.

\begin{figure}[t]
\scalebox{1.18}
{\plotone{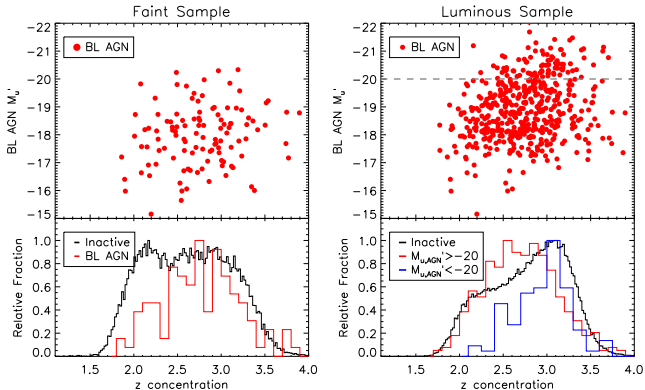}}
\figcaption{At top, the host-subtracted AGN $u$-band absolute
  magnitude with $z$-band concentration for both the faint
  ($M_r'<-19$, $0.01<z<0.05$) and luminous ($M_r'<-20.8$,
  $0.01<z<0.11$) samples.  The bottom panels show histograms of the
  $C_z$ distributions for both inactive galaxies and broad-line AGNs
  in both samples.  The distribution of concentration among the most
  luminous ($M_{u,\rm AGN}'<-20$) AGNs peaks at higher values: we flag
  these high-luminosity AGNs in the subsequent discussion because
  their concentration might be biased by AGN contamination.  Note that
  the different concentration distributions between the two inactive
  galaxy samples is caused by the luminous sample containing a higher
  proportion of high-mass spheroids than the faint sample.
\label{fig:agnzconc}}
\end{figure}

Using Equation \ref{eq:cdeltamag} for AGN/host decomposition has two
important assumptions.  First, we assume that $C_z$ is not
contaminated by the broad-line AGN.  We originally chose $C_z$ with
this in mind: $z$ band light is the least affected by the (typically
blue) AGN, and $R_{50}$ and $R_{90}$ are large radii well beyond the
point-source AGN.  Figure \ref{fig:agnzconc} directly tests if the AGN
affects $C_z$ by plotting AGN luminosity against $z$-band
concentration and comparing the distributions of $C_z$ among
broad-line AGNs and inactive galaxies.  In general, both the faint and
luminous AGN samples span a wide range of concentrations.  However the
most luminous ($M_{u,\rm AGN}'<-20$) AGNs are typically more
concentrated: this may be evidence that very luminous AGNs contaminate
the concentration measurement, or it may be that more luminous AGNs
prefer more bulge-like hosts \citep{bah97,cis11,koc12}.  In case it is
the result of a bias, we flag these high-luminosity AGNs when
discussing any connections between AGN strength and host galaxy
properties.  Meanwhile we conclude that our assumption that $C_z$ is
unaffected by the AGN remains valid for the bulk of broad-line AGNs
with $M_{u,\rm AGN}'>-20$.

\begin{figure}[t]
\scalebox{1.18}
{\plotone{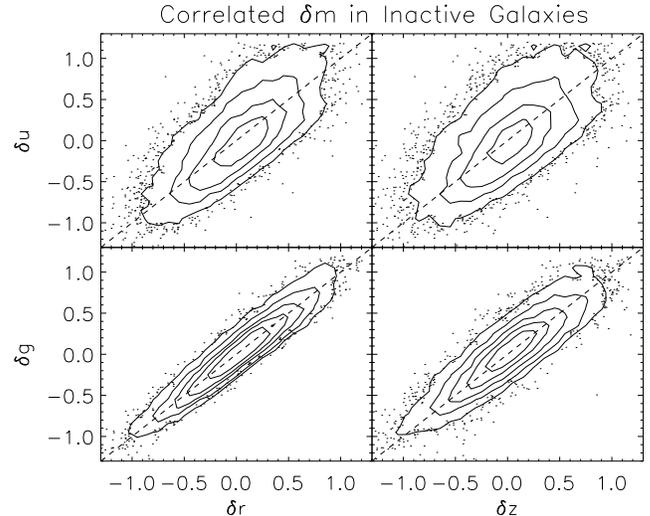}}
\figcaption{Comparisons of measured ${\delta}m$ (from Equation
  \ref{eq:cdeltamag}) for inactive galaxies in $ugrz$ filters.  Values
  of ${\delta}m$ in different filters are strongly correlated near the
  dashed one-to-one line.
\label{fig:cdeltavs}}
\end{figure}

The second assumption is that all inactive galaxies have ${\delta}m
\sim 0$ with some random scatter $\sigma_{{\delta}m}$, independent of
the presence of an AGN.  There is evidence that this scatter is not
due to measurement error: as Figure \ref{fig:cdeltavs} shows, inactive
galaxies have values of ${\delta}m$ that are correlated across the
$ugriz$ filters.  In other words, an inactive galaxy with ${\delta}m
\sim 1$ in the $u$ band will also tend to have ${\delta}m \sim 1$ in
$g, r, i, z$.  (This effectively means that the ${\delta}m=0$
assumption causes a much smaller scatter in color than in luminosity
or stellar mass: see Section 3.4.)  We tested if, in addition to
$C_z$, ${\delta}m$ was connected to galaxy properties like color or
\citet{ser68} index.  However we found no additional correlation and
were unable to find the physical basis for the small scatter of
inactive galaxies about ${\delta}m=0$.  Defining ${\delta}m$ using
color and S\'{e}rsic index instead of $C_z$ also proved ineffective at
reducing the scatter.  Instead $\sigma_{{\delta}m}$ is treated as a
random error, and we investigate its effects in Section 3.4.

Inactive galaxies with ${\delta}z>1$ (in the upper left of the right
panels in Figure \ref{fig:cdeltamag}) are potentially interesting
because they have similar relationships between inner and outer
magnitudes to bright AGNs.  However after visually inspecting their
images and spectra we determined that these galaxies probably have
nuclear starbursts and are not some class of misclassified broad-line
AGNs.

\subsection{Point Source Inner and Outer Aperture Magnitudes}

Most of the light from a point source is detected in the inner
aperture, but the PSF of the SDSS causes light from point sources to
``leak'' into the outer aperture as well.  Stars can be used to model
the relationship between inner and outer aperture magnitudes for
AGN-only light ($m_{in,AGN}$ and $m_{out,AGN}$), since both are point
sources.  We select a sample of 362 stars with $r<16$ and
$-0.2<r-z<-0.1$.  The brightness ensures their photometry is
well-measured, and the color cut makes them similar to the colors of
bright AGNs.

\begin{figure}[t]
\scalebox{1.18}
{\plotone{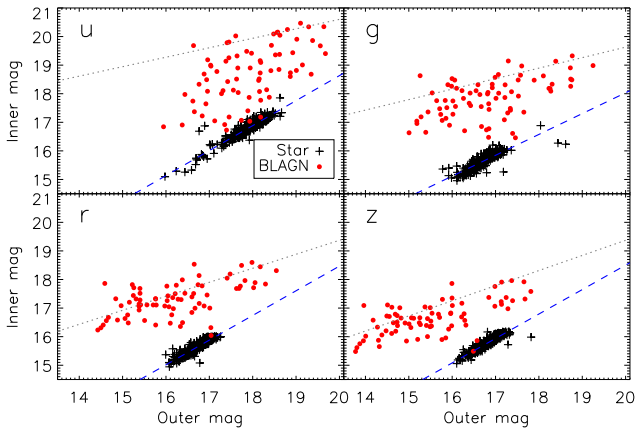}}
\figcaption{Inner and outer magnitudes in the $ugrz$ filters for stars
  (black crosses) and broad-line AGNs at $0.05<z<0.06$ (red circles).
  The blue dashed line in each panel shows the best-fit line for
  point-source stars, while the dotted gray line shows the best-fit
  line for $0.05<z<0.06$ inactive galaxies (not shown).  Broad-line
  AGNs tend to scatter between the two lines, with the brightest AGNs
  behaving like point sources (especially in $u$), and more modest
  AGNs behaving like inactive galaxies (especially in $r$ and $z$).
\label{fig:starinnerouter}}
\end{figure}

Figure \ref{fig:starinnerouter} shows the inner and outer magnitudes
for the sample of 362 stars.  These point sources have a well-defined
relationship between inner and outer aperture magnitude, and the
best-fit line in each filter describes the relationship between inner
and outer magnitude for the AGN-only light:
\begin{equation}
  m_{in,AGN} = G + H m_{out,AGN} \label{eq:agnmag}.
\end{equation}
Figure \ref{fig:starinnerouter} shows that the brightest AGNs behave
like point sources in blue light and lie on this line, while in red
light AGNs behave more like inactive galaxies.  Because all the images
are convolved to 2\arcsec~before measuring the aperture photometry,
$m_{in,AGN} \approx 1 + m_{out,AGN}$ in all five filters.

\subsection{The Equations for Decomposing AGN and Galaxy Light}

Given the derived relationships between inner and outer magnitudes for
galaxy-only (Equations \ref{eq:deltamag} and \ref{eq:cdeltamag}) and
AGN-only (Equation \ref{eq:agnmag}) light, we can solve for the
unknown quantities $m_{in,AGN}$, $m_{out,AGN}$, $m_{in,GAL}$, and
$m_{out,GAL}$.  First, the total outer and inner magnitudes (both
measured quantities) are simply the sum of the flux contributions from
the AGN and the galaxy:
\begin{equation}
  f_{in} = f_{in,AGN} + f_{in,GAL} \label{eq:inner}
\end{equation}
\begin{equation}
  f_{out} = f_{out,AGN} + f_{out,GAL} \label{eq:outer}
\end{equation}

The relationship between flux and AB magnitude is defined by $m \equiv
-2.5 \log(f_\nu)-48.6$.  Combining Equations \ref{eq:deltamag} and
\ref{eq:cdeltamag} and converting to flux results in:
\begin{equation}
  f_{in,GAL} = X \cdot f_{out,GAL}^B, \label{eq:galfl}
\end{equation}
with $X = 10^{-0.4(A + 48.6 - 48.6B - C - D C_z - E C_z^2 - F C_z^3)}$.

Equation \ref{eq:agnmag} is similarly converted from magnitude to
flux:
\begin{equation}
  f_{in,AGN} = Q \cdot f_{out,AGN}^H, \label{eq:agnfl}
\end{equation}
with $Q = 10^{-0.4(G + 48.6 - 48.6H)}$.

There are now four equations (Equations \ref{eq:inner},
\ref{eq:outer}, \ref{eq:galfl}, and \ref{eq:agnfl}) that describe the
relationships between inner and outer magnitudes for each of the
broad-line AGN and galaxy contributions.  The quantities $f_{in}$,
$f_{out}$, and $C_z$ are measured, and the coefficients $A$, $B$, $C$,
$D$, $E$, $F$, $G$, and $H$ come from our line fits in each filter and
redshift bin (with three example filters and redshift bins shown in
Figures \ref{fig:innerouter}, \ref{fig:deltamag}, and
\ref{fig:cdeltamag}).  Solving these four equations for $f_{in,AGN}$
results in:

\begin{equation}
  f_{in,AGN} = f_{in}-X [f_{out}-(f_{in,AGN}/Q)^{1/H}]^B
    \label{eq:solution}
\end{equation}
(with the constants $X$ and $Q$ given in equations \ref{eq:galfl} and
\ref{eq:agnfl}).  We numerically solve this equation using a bisector
root-finding method.  After solving for $f_{in,AGN}$, the other three
unknowns can be computed from Equations \ref{eq:inner},
\ref{eq:outer}, and \ref{eq:agnfl}.  This process is repeated to
derive AGN-subtracted host galaxy magnitudes in all five bands in each
of the ten redshift bins.  Table \ref{tbl:catalog} shows the corrected
magnitudes and derived stellar masses for the 820 broad-line AGNs in
the luminous and faint samples.

\subsection{Error Analysis}

\begin{figure*}[t]
\begin{center}
  {\plotone{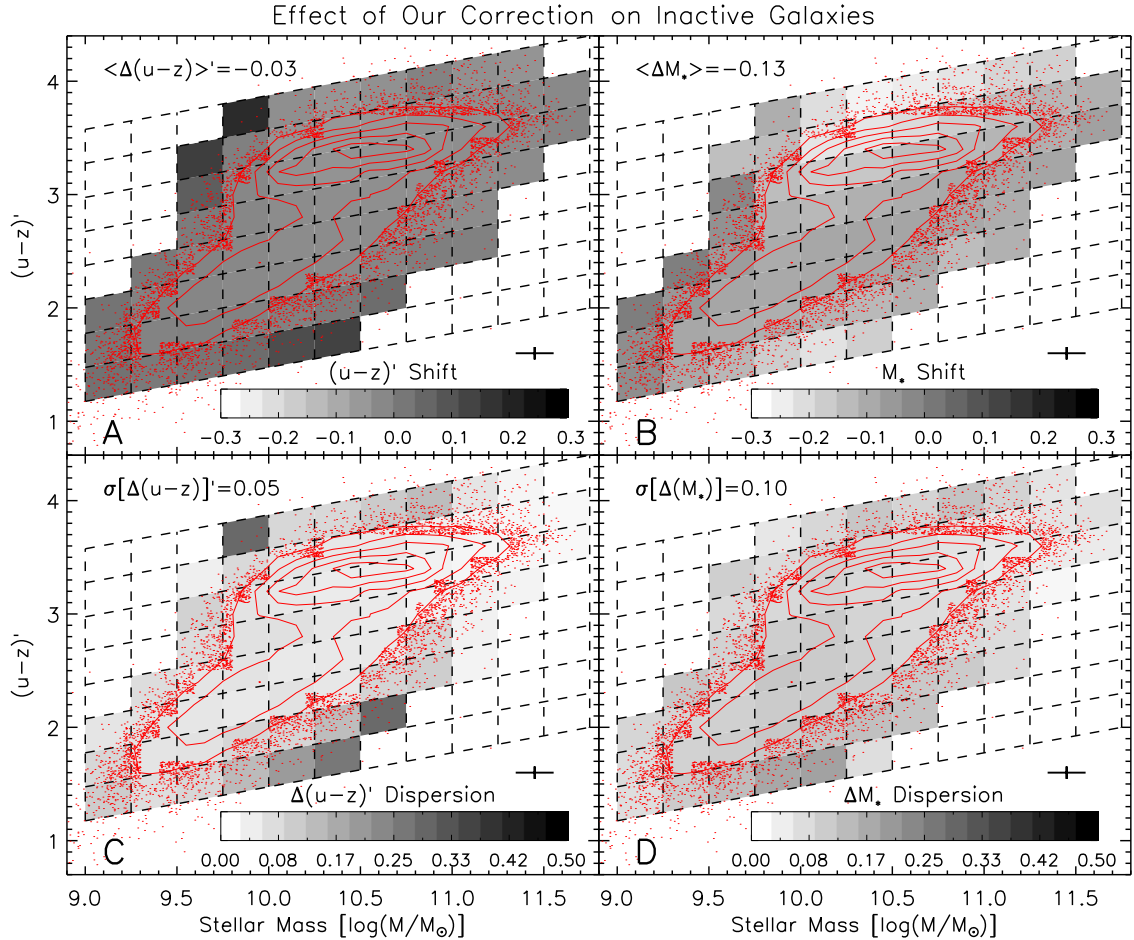}}
\end{center}
\figcaption{Bins in color and stellar mass showing the effects of our
  AGN/host decomposition method applied to inactive galaxies.  The top
  panels (A and B) show the average shifts in $(u-z)'$ color and
  stellar mass, and the bottom panels (C and D) show the dispersion of
  the changes in color and mass.  The text in the upper left of each
  panel also gives the mean shift or dispersion over all galaxies.
  Outside of bins with small numbers of galaxies, there are no regions
  of color-mass space with unusually large shifts or dispersions in
  corrected color and mass.
\label{fig:galerrors}}
\end{figure*}

It is necessary to ensure that our AGN/host decomposition method does
not introduce errors which bias the resultant host galaxy colors and
stellar masses.  One assumption of our method is that galaxies have
${\delta}m=0$: Figure \ref{fig:cdeltamag} demonstrates that this is
true on average, but the ${\delta}m$ values of individual galaxies
scatter about this value.  Similarly the true inner and outer
magnitudes of AGN-subtracted host galaxies may have nonzero values of
${\delta}m$, and our correction to ${\delta}m=0$ may introduce
systematic errors in their resultant colors and masses.

We investigate the errors of the ${\delta}m=0$ assumption by applying
the decomposition method to inactive galaxies.  Figure
\ref{fig:galerrors} shows the changes in $(u-z)'$ color and $M_*$
introduced by ``correcting'' inactive galaxies to ${\delta}m=0$.  The
dispersions in these shifts represent the errors in color and mass
from correcting AGN host galaxies with nonzero ${\delta}m$.  Errors in
color are small because ${\delta}u$ and ${\delta}z$ are correlated
(see Figure \ref{fig:cdeltavs}).  The dispersion in stellar mass is
slightly larger, as indicated by the error bar in the bottom right of
each panel, but mass shifts and dispersions are small among blue cloud
and green valley galaxies (the typical hosts of broad-line AGNs).
From Figure \ref{fig:galerrors}, we do not expect the ${\delta}m=0$
assumption to bias the observed colors and masses of broad-line AGNs.

If our other assumption is incorrect and AGNs contaminate $z$-band
concentration, this could also introduce significant errors in the
AGN/host decomposition method.  Figure \ref{fig:agnzconc} shows that
the most luminous AGNs ($M_{u,\rm AGN}'<-20$) may bias $C_z$, and such
systems may have larger systematic or random errors than those
estimated in Figure \ref{fig:galerrors}.  For this reason we eliminate
the most luminous AGNs from the discussion and conclusions below.
However the low and moderate luminosity broad-line AGNs that make up
the bulk of our sample do not exhibit biased $C_z$.  For these objects
the assumption that AGNs do not affect $z$-band concentration is
unlikely to cause errors in the AGN/host decomposition.

\begin{figure*}[t]
\begin{center}
  {\plotone{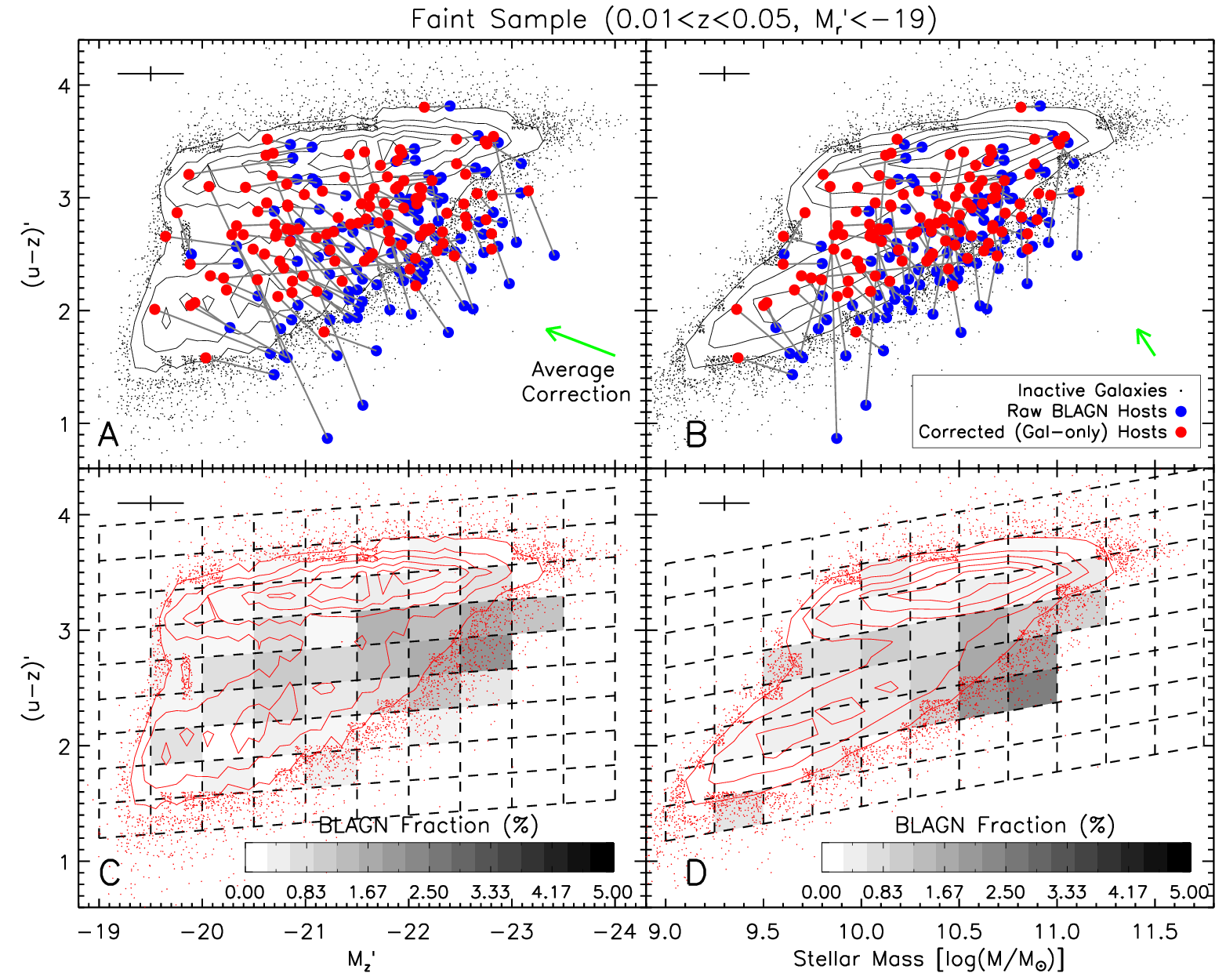}}
\end{center}
\figcaption{The color-luminosity and color-mass diagrams for inactive
  galaxies and broad-line AGN hosts in the faint sample ($M_r'<-19$
  and $0.01<z<0.05$).  The top panels (A and B) shows total $(u-z)'$
  color vs. luminosity and stellar mass for both uncorrected
  broad-line AGNs (blue filled circles) and AGN-subtracted host
  galaxies (red filled circles), with lines connecting the two
  measurements of each AGN.  The mean AGN subtraction vectors are
  shown by the green arrows.  The bottom panels (C and D) show the
  broad-line AGN frequency in bins of color and luminosity or color
  and mass, using the corrected AGN host properties.  The typical
  scatter of our AGN/host decomposition method (see Section 3.4) is
  shown by error bars in the upper left of each panel, and contours
  and points represent the inactive galaxy populations.  Star-forming
  (blue cloud and green valley) galaxies are the most common hosts of
  broad-line AGNs.
\label{fig:cmdfaint}}
\end{figure*}

\begin{figure*}[t]
\begin{center}
  {\plotone{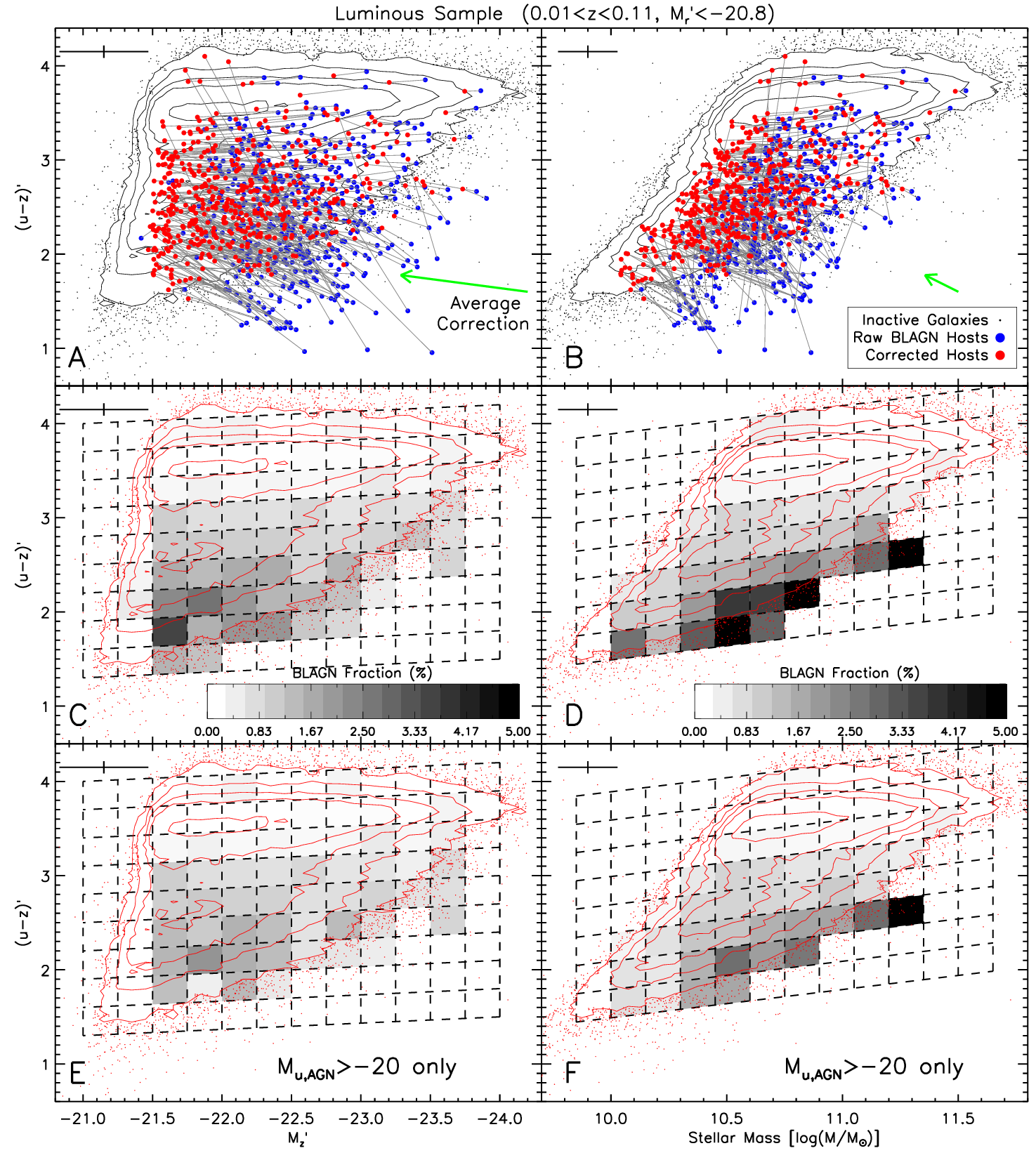}}
\end{center}
\figcaption{The color-luminosity and color-mass diagrams for inactive
  galaxies and broad-line AGN hosts in the luminous sample
  ($M_r'<-20.8$ and $0.01<z<0.11$).  The top panels (A and B) show
  uncorrected (blue) and corrected (red) total $(u-z)'$ color
  vs. luminosity and stellar mass for broad-line AGN hosts, and the
  middle panels (C and D) show the AGN frequency in bins of color and
  luminosity or color and mass.  The bottom panels (E and F) include
  only AGN hosts with $M_{u,\rm AGN}'>-20$: that is, AGNs with
  potentially biased $C_z$ are removed.  Error bars in the upper left
  of each panel represent the typical scatter introduced by AGN/host
  decomposition (see Section 3.4), and inactive galaxies are shown by
  contours and points.  As in the faint sample, broad-line AGNs are
  most common in the bluest galaxies of a given stellar mass.
\label{fig:cmdlumin}}
\end{figure*}

\section{The Host Galaxies of Broad-Line AGNs}

The corrected galaxy-only photometry of the broad-line AGNs can be
used to determine if certain types of host galaxies are more likely to
exhibit broad-line AGNs.  After computing the AGN-subtracted
photometry, we K-correct to $z=0.05$ and infer the host galaxy stellar
mass using Figure \ref{fig:masslight}.  Figure \ref{fig:cmdfaint}
shows total $(u-z)'$ color vs. luminosity and stellar mass for
broad-line AGNs in the faint sample, with contours showing the
inactive galaxies for comparison.  Figure \ref{fig:cmdlumin} similarly
shows broad-line AGNs and inactive galaxies in the luminous sample.

The top panels of each figure show that broad-line AGNs typically
become redder, dimmer, and less massive after recovering the
uncontaminated galaxy light.  A few AGN hosts become so much redder
without the blue point-source AGN that their masses increase, due to
the higher mass-to-light ratios of similarly red inactive galaxies.
Because we independently correct each filter, there are also a few AGN
hosts that become bluer, presumably because they contain a reddened
AGN.

To determine if broad-line AGNs are more likely to occur in certain
types of galaxies, we measure the fraction of galaxies containing an
AGN in bins across the color-mass diagram, as shown in the bottom
panels of Figures \ref{fig:cmdfaint} and \ref{fig:cmdlumin}.  Bins are
tilted (with slopes of $\Delta(u-z)' / \Delta{M_z'}=-0.07$ and
$\Delta(u-z)' / \Delta[\log(M_*/M_\odot)]=0.3$) to be parallel to the
red sequence.  We count the number of broad-line AGNs with corrected
galaxy-only photometry in each bin, then divide by the total number of
galaxies (inactive plus active) in that bin to determine the AGN
fraction.  Table \ref{tbl:agnfrac} gives the broad-line AGN fraction
in each bin of color and mass, as well as the AGN fraction over the
entire mass slice.

In both samples, galaxies on the red sequence are least likely to host
a broad-line AGN.  Instead AGNs are most likely to be in the bluest
host galaxies of a given stellar mass.  We further explore what the
preference for star-forming hosts means for the AGN-SF connection and
AGN feedback in Section 5, but first test to ensure that our results
are not caused by selection effects.

\subsection{Selection Effects}

While broad-line AGNs are the brightest and most rapidly accreting
AGNs \citep[e.g.,][]{kol06,tru09}, many of the weakest broad-line AGNs
have significant galaxy contribution to their continua.  Dim AGNs
could be more difficult to detect in bright galaxies, with the broad
lines diluted by host galaxy starlight.  More distant galaxies also
contain more starlight within the 3\arcsec~spectroscopic fiber
aperture and could similarly dilute the broad emission lines of AGNs.

\begin{figure*}[t]
\begin{center}
  {\plotone{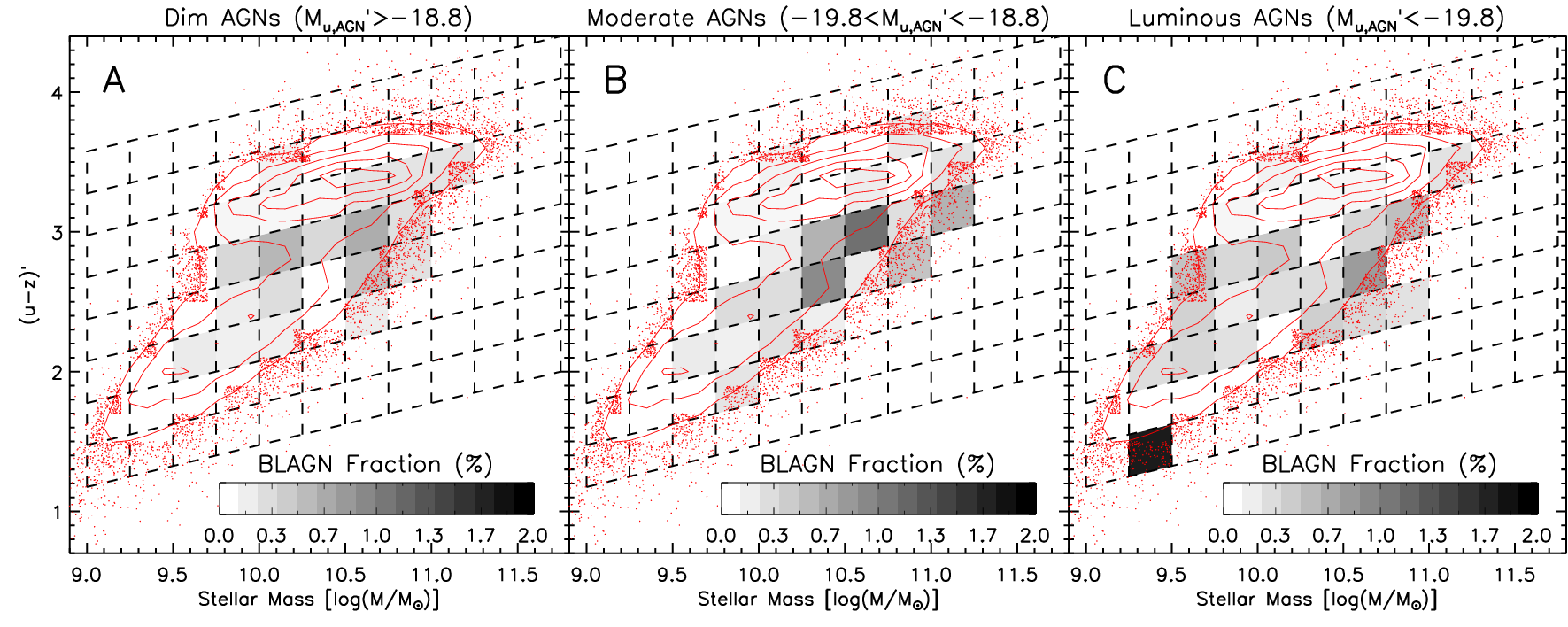}}
\end{center}
\figcaption{The frequency of dim ($M_{u,\rm AGN}'>-18.8$), moderate
  ($-19.8<M_{u,{\rm AGN}}'<-18.8$), and luminous ($M_{u,{\rm
      AGN}}'<-19.8$) broad-line AGNs across the color-mass diagram,
  drawn from the faint sample ($M_r'<-19$, $0.01<z<0.05$).  The red
  contours and points show the inactive galaxy population.  As in
  Figures \ref{fig:cmdfaint} and \ref{fig:cmdlumin}, the AGN fraction
  is calculated in each bin using the corrected galaxy-only
  photometry.  Dim AGNs are no less likely to be in high-mass galaxies
  than strong AGNs, suggesting that there are no selection effects
  against identifying dim AGNs in bright galaxies.
\label{fig:cmdqsostr}}
\end{figure*}

We test the potential selection bias against dim AGNs in luminous host
galaxies by computing the AGN fraction across the color-mass diagram
for dim, moderate, and luminous AGNs, as shown in Figure
\ref{fig:cmdqsostr}.  AGN strength is quantified using the AGN-only
$u$-band absolute magnitude, with $M_{u,{\rm AGN}}$ divisions chosen
to put the same number of AGNs in each set.  AGNs were drawn from the
faint sample because it includes a large range in stellar mass.  From
Figure \ref{fig:cmdqsostr} it is clear that dim AGNs are not found
less frequently in massive, luminous galaxies, and so this potential
selection effect does not occur in our sample.

\begin{figure}[t]
\scalebox{1.15}
{\plotone{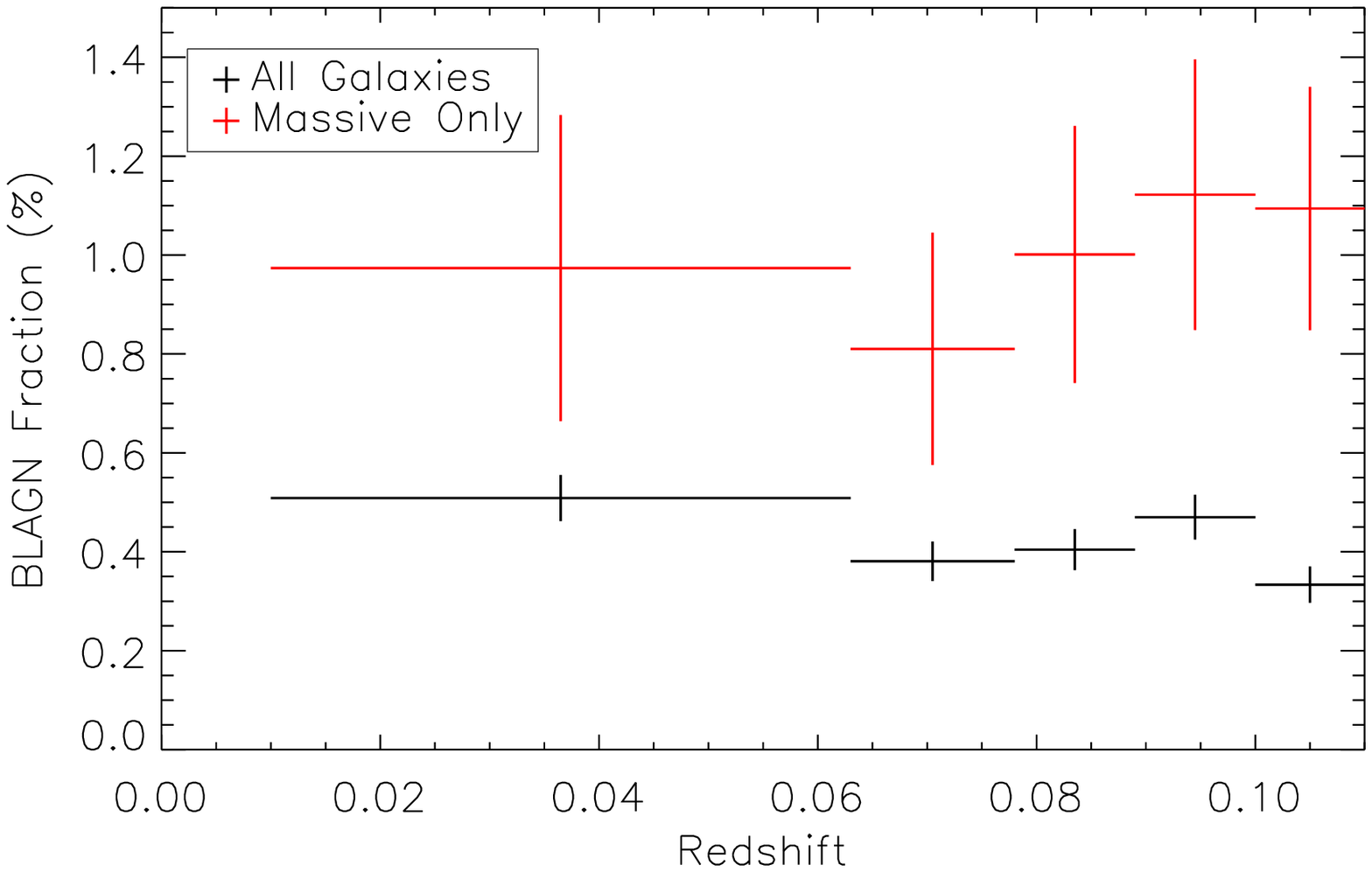}} 
\figcaption{The fraction of galaxies from the luminous sample
  containing a broad-line AGN with redshift.  Black lines show the
  population of all galaxies, while red lines show only with
  $\log(M_*/M_\odot) > 9.5+0.5(u-z)'$ (this corresponds to the most
  massive 10\% of galaxies on the blue cloud and green valley).  Each
  redshift bin contains the same number of inactive galaxies, and the
  vertical error bars indicate the 1$\sigma$ error assuming Poissonian
  statistics.  There is no evidence for fewer broad-line AGNs at
  higher redshift, suggesting that our sample is quite complete.
\label{fig:qsofrac_redshift}}
\end{figure}

Figure \ref{fig:qsofrac_redshift} similarly tests the potential
selection effect of higher distance biasing against AGN
identification.  The luminous sample is most appropriate for this test
because it includes luminosity-limited AGNs over the full redshift
range $0.01<z<0.11$.  In addition to measuring the broad-line AGN
fraction over all galaxies in each redshift bin, a set of ``massive''
galaxies is also defined with $\log(M_*/M_\odot) > 9.5+0.5(u-z)'$.
This mass limit corresponds to the most massive 10\% of galaxies along
the blue cloud and green valley.

Over the limited $0.01<z<0.11$ redshift range of our sample, there is
no evidence for a decline in the fraction of galaxies hosting a
broad-line AGN at higher distance.  For this reason we do not expect
distant galaxies to bias against AGN identification through starlight
dilution in the spectroscopic aperture.

\section{Discussion}

In this section we discuss the consequences of AGN hosts with young
stellar populations for the coevolving growth of galaxies and
supermassive black holes.

\subsection{AGNs and Star Formation}

A major finding of this study is that broad-line AGNs tend to lie in
hosts with young stellar populations, and avoid red and quiescent host
galaxies.  The simplest explanation for this is if the same gas that
fuels rapidly accreting, unobscured nuclear activity also rapidly
formed stars in the recent past.  Figure \ref{fig:cmdqsostr} suggests
that the more powerful the broad-line AGN, the bluer the host galaxy:
more rapidly accreting AGNs may have more powerful recent star
formation in their host galaxy.  Once again, this suggests that star
formation and nuclear activity are fueled by the same material.

\begin{figure*}[t]
\begin{center}
  {\plotone{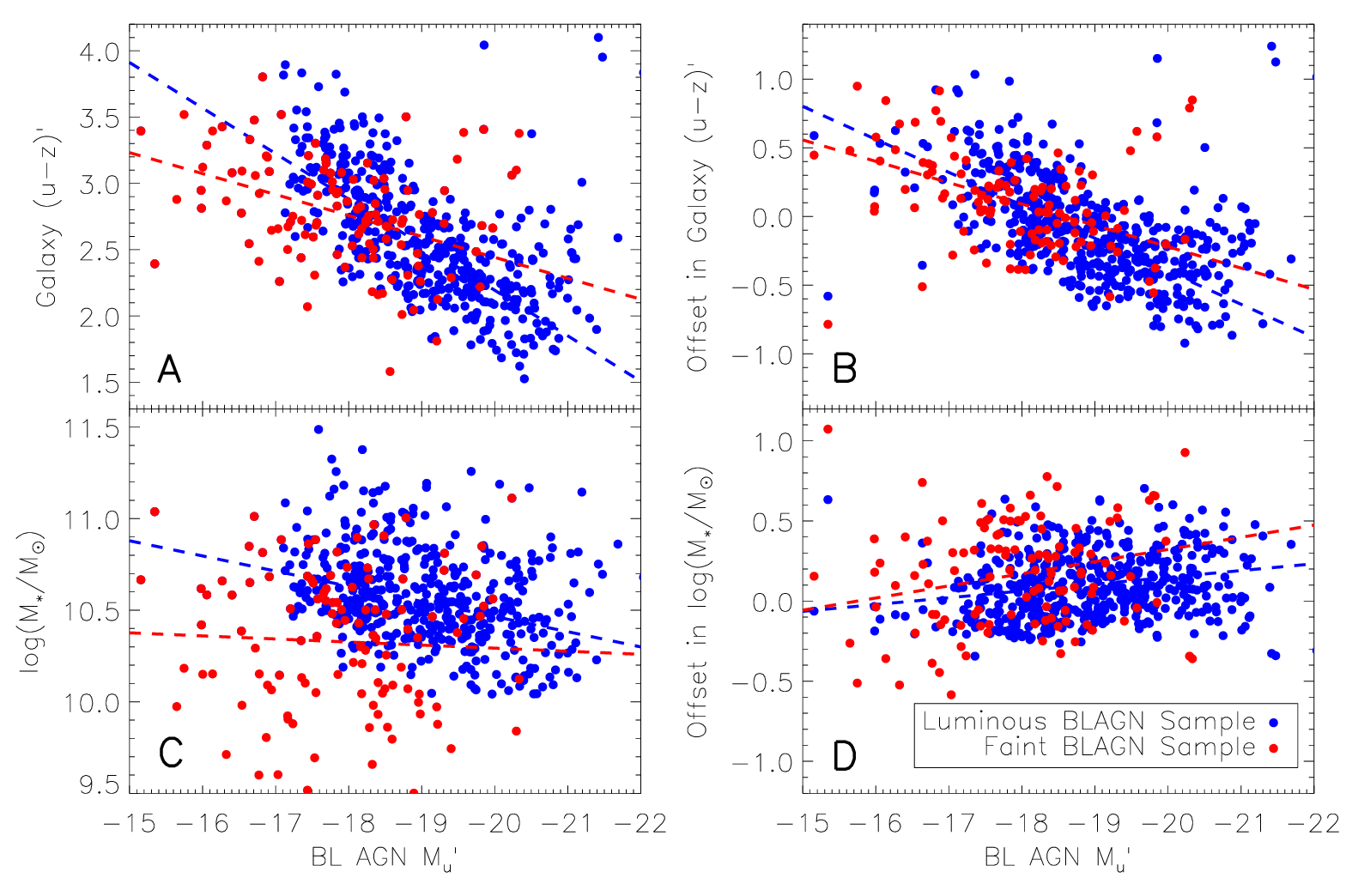}}
\end{center}
\figcaption{Corrected host galaxy $(u-z)'$ color and stellar mass
  vs. AGN luminosity $M_u'$.  Panel B (top right) shows the color
  offsets of broad-line AGNs, defined as the difference between the
  corrected AGN host color and the median color of inactive galaxies
  with similar mass.  Similarly Panel D (bottom right) shows the
  offset in BL AGN host stellar mass from the median stellar mass of
  inactive galaxies with similar color.  Red points show the faint
  broad-line AGN sample ($M_r'<-19$, $0.01<z<0.05$), while blue points
  show the luminous sample ($M_r'<-20.8$, $0.01<z<0.11$).  Dashed
  lines indicate the best-fit line for AGNs with $M_{u,\rm AGN}'>-20$:
  the fit excludes the most luminous AGNs because they may have biased
  $C_z$ measurements (see Section 3.1 and Figure \ref{fig:agnzconc}).
  Galaxies hosting more powerful AGNs typically have younger stellar
  populations, suggesting that the same gas needed to fuel a
  broad-line AGN also drives recent star formation activity.
\label{fig:qsostr}}
\end{figure*}

Figure \ref{fig:qsostr} directly compares the broad-line AGN
luminosity to the color and stellar mass of the host galaxy.  AGN
luminosity is quantified by AGN-only $M_u'$.  In addition to corrected
color and mass for broad-line AGN hosts, Figure \ref{fig:qsostr} also
shows the offsets in each of these quantities from inactive galaxies
of similar (within 0.1~dex) mass or color.  These offsets are useful
to remove the degeneracy between a galaxy's mass and its color.  For
example, the apparent anti-correlation between stellar mass and AGN
luminosity in Panel C is a result of the degeneracy between mass and
color.  Comparing AGNs and inactive galaxies of similar color, Panel D
shows no significant trend between stellar mass offset and AGN
luminosity.

Panel B in Figure \ref{fig:qsostr} shows that more luminous AGNs have
bluer host galaxies in both the luminous and faint samples.  This may
represent a physical connection between AGN luminosity and the host
galaxy's recent star formation history.  AGN luminosity probably
translates to accretion rate, and so the relationship between AGN
$M_u'$ and host color indicates that AGN accretion rate is correlated
with the number of recently formed stars.  These observations favor
scenarios with similar fuel sources for both star formation and
nuclear activity \citep[e.g.,][]{sal07,sil09}, with little or no time
delay between the processes fueling each.

This interpretation is robust even given the limitations of our
AGN/galaxy decomposition method.  Recall the assumption that AGN host
galaxies have the same light distributions as inactive galaxies
(quantified by $\delta{m}=0$).  Given the $M-\sigma$ relation
\citep[e.g.][]{park12} and requirement of a bulge to host an AGN, it
is unlikely that broad-line AGNs have hosts with instrinsic
$\delta{m}<0$.  On the other hand, the unusual blue colors of
broad-line AGN hosts might indicate nuclear starbursts and intrinsic
$\delta{u}>0$.  Applying our decomposition method to an AGN host with
a nuclear starburst would over-subtract the blue light, resulting in a
redder corrected galaxy $(u-z)'$ color.  With the presence of a
nuclear starburst it is possible that the galaxy colors of AGN hosts
may be even bluer than we observe.  Similarly the connection between
AGN luminosity and blue host color in Figure \ref{fig:qsostr} would
steepen if these blue hosts contain nuclear starbursts and unusually
concentrated blue light.

Another potential concern is that the broad-line AGN host galaxies
appear blue due to extended scattered light from the AGN.
\citet{zak06} noted that in extreme cases Type 2 quasars with
$M_{B,AGN}<-24$ have scattered light on $>$1~kpc scales that
dominates over the host galaxy starlight.  Measuring extended
scattered light for our sample is beyond the scope of this work, given
the requirement for long-slit spectropolarimetry on a large number of
broad-line AGNs with varying luminosities.  However we note that our
broad-line AGNs are several (4-7) magnitudes fainter than the
\citet{zak06} sample, and so we assume the amount of extended
scattered light in this work to be neglible.


\subsection{The Different Hosts of Luminous Broad-Line and Weak Narrow-Line AGNs}

Our large sample of broad-line AGNs from the SDSS shows a strong
preference for blue host galaxies.  This matches earlier studies with
small broad-line AGN samples, which found a similar preference for
hosts with young stellar populations \citep{bah97,jah04a,jah04b}.  The
most $\OIII$-luminous narrow-line AGNs also prefer star-forming host
galaxies \citep{kau03b,sil09}.  These classes of luminous AGNs all
represent ``feast mode'' rapid accretion with high Eddington ratios
($\lambda_{Edd}$): \citet{kau09} estimate that the most
$\OIII$-luminous narrow-line AGNs have $\lambda_{Edd} \sim 0.1$, and
broad-line AGNs also have Eddington ratios of 1-100\%
\citep{kol06,tru09}.  In future work we will directly measure and
compare $\lambda_{Edd}$ in our broad-line AGN sample, but for now we
simply assume that they are rapidly accreting AGNs with
$\lambda_{Edd}>0.01$.

The blue hosts of our broad-line AGNs are in contrast to the hosts of
fainter narrow-line AGNs, which instead prefer green valley galaxies
\citep{nan07, sal07, geo08, sil08, gab09, schaw09, hic09, koc09} or
all galaxy types equally \citep{xue10}.  The host-dominated AGNs of
these previous studies are, by construction, fainter than broad-line
AGNs, due to either obscuration or lower accretion rates
($\lambda_{Edd}<0.01$).  Different host galaxies for broad-line AGNs
and weaker host-dominated AGNs are incompatible with the historical
AGN unified model \citep{ant93}.  In its simplest version, the unified
model uses only geometrical orientation to explain the different
observed properties of broad-line and host-dominated AGNs.  Both
broad-line and host-dominated AGNs would then be observed in the same
host galaxy types, as has been observed among intermediate-luminosity
AGNs at $z \sim 1$ \citep{amm11}.  Our observations instead favor AGN
unified models where broad-line and faint narrow-line AGNs have
physically different accretion properties \citep{ho08,tru11,ant11}.
Broad-line AGNs might be in star-forming galaxies because such
galaxies have the most abundant or efficient fuel supply for the SMBH,
while other galaxy types can fuel only ``famine-mode'' host-dominated
AGNs with low accretion rates \citep{kau09,schaw10}.

\begin{figure}
\scalebox{1.2}
{\plotone{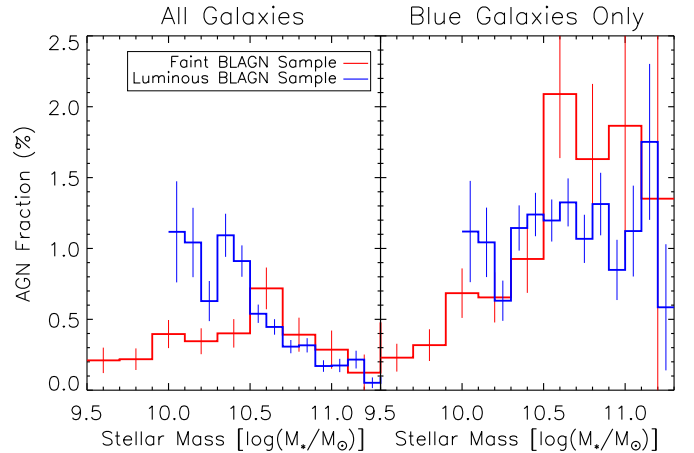}}
\figcaption{The percentage of AGNs found in hosts of varied stellar
  masses, for both the faint (red lines) and luminous (blue lines)
  broad-line AGN samples.  Vertical lines give the Poissonian errors
  in each bin.  At left, the luminous sample suggests that AGNs are
  more common in low-mass galaxies, but this is driven purely by the
  preference for blue hosts.  The right panel instead shows the AGN
  fraction in blue galaxies with $(u-z)'<0.3\log(M_*/M_\odot)$.  Here
  AGNs in the luminous sample have equivalent likelihoods to be in
  hosts of any stellar mass, in agreement with \citet{aird12}.
  Meanwhile, the faint sample may show a preference for AGNs in
  high-mass hosts, but this is not significant and is perhaps the
  result of small number statistics or incompleteness at
  $\log(M_*/M_\odot)<10$.
\label{fig:massbins}}
\end{figure}

\citet{kau03b} used a large sample of host-dominated AGNs to show that
there may be a minimum stellar mass of $\log(M_*/M_\odot)>10.5$
required for a galaxy to host an AGN.  Recently, however,
\citet{aird12} argued that the apparent preference for massive AGN
hosts is caused by selection effects, and there is instead a universal
Eddington-ratio distribution of AGNs in galaxies of any stellar mass.
Figure \ref{fig:massbins} shows the percentage of broad-line AGNs with
host stellar mass for both the luminous and faint samples.  Given the
narrow distribution of high Eddington ratios for broad-line AGNs
\citep[$0.01<L/L_{Edd}<1$, e.g.,][]{kol06,tru09}, our AGN sample is
complete to $\log(M_*/M_\odot) \sim 10$ \citep{kelly12}.  Thus the
luminous broad-line AGN sample (also limited by
$\log(M_*/M_\odot)>10$) provides a direct test of the AGN dependence
on host stellar mass.  After controlling for galaxy color (right panel
of Figure \ref{fig:massbins}), we find a flat AGN fraction with
stellar mass.  This suggests that there is no stellar mass threshold
for broad-line AGNs, in agreement with \citet{aird12}.

\subsection{Consequences for AGN Feedback}

\begin{figure*}[t]
\begin{center}
  {\plotone{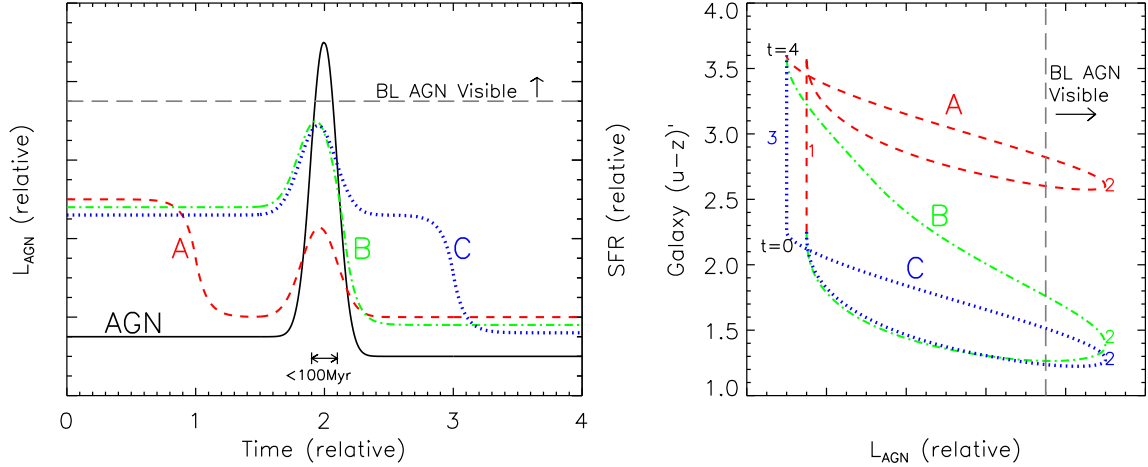}}
\end{center}
\figcaption{A phenomenological model outlining three scenarios
  connecting AGN activity and the quenching of star formation.  The
  broad-line AGN phase is parameterized as a Gaussian function with a
  width of less than $100$~Myr \citep{mar04}, and the three curves A,
  B, and C represent star formation rates with quenching occurring
  before, during, and after the broad-line AGN phase.  The left panel
  shows the evolution of the AGN and SFR functions with time, and the
  right panel shows the expected galaxy colors vs. AGN luminosity for
  the same curves.  Numerals along the curves in the right panel
  correspond to times from the left panel ($t=1$ is not shown for
  curves B or C because it is identical to $t=0$, just as $t=3$ is the
  same as $t=4$ for curves A and B).  Our observations rule out
  scenario A, but allow for quenching either during or after the
  broad-line AGN phase as shown by curves B and C.
\label{fig:quenchcartoon}}
\end{figure*}

Theoretical simulations invoke feedback from luminous AGNs to rapidly
quench star formation and transform their host galaxy colors from blue
to red \citep{sil98,fab02,dim05}.  Several authors have claimed
observational evidence for this scenario, with host-dominated AGNs
apparently preferring recently-quenched green valley galaxies
\citep{nan07,geo08,schaw09,hic09,koc09}.  Since radiative-mode
feedback scales with AGN strength, the broad-line AGNs in our sample
would be expected to have an even greater effect in quenching star
formation.  However, we find precisely the opposite effect: more
powerful AGNs seem to lie in bluer galaxies.  If quasar winds cause
significant feedback, then their galaxy-wide effects are not visible
until after the broad-line AGN disappears.

Figure \ref{fig:quenchcartoon} outlines three scenarios connecting
AGNs and galaxy quenching: scenario A has star formation quenched
before the broad-line AGN phase, scenario B quenches during the
broad-line AGN peak, and scenario C has star formation quenched well
after the broad-line AGN phase.  The increase in star formation during
the AGN peak in all scenarios is motivated by our observed connection
between AGN luminosity and blue host galaxy color.  The broad-line AGN
phase occurs only during the most luminous AGN activity, which means
that our study probes only the region limited by the dashed ``BL AGN
Visible'' lines in each panel.

Scenario A is immediately ruled out by the observed absence of
broad-line AGNs among quenched red galaxies.  However our observations
cannot distinguish between quenching during or after the broad-line
AGN phase (scenarios B and C).  The broad-line AGN lifetime is less
than $\sim$100~Myr \citep{mar04}, which is roughly the same amount of
time it takes for galaxy colors to change from blue to green or red
after star formation quenches \citep[e.g.,][]{bru03}.  Even if
quenching occurs during the AGN peak, a change in galaxy colors would
not be observed until after the broad-line AGN fades.  This is evident
in the scenario B curve in the right panel of Figure
\ref{fig:quenchcartoon}.  So while our observations exclude AGN
quenching scenarios that occur before the broad-line AGN phase, only
observations of fading AGNs can distinguish between simultaneous or
delayed AGN quenching.

\section{Summary}

An aperture photometry method is used to disentangle the light of
broad-line AGNs and their host galaxies at $0.01<z<0.11$.  Based on
the corrected, host-only photometry, broad-line AGNs are distributed
throughout the blue cloud and green valley but are very rare among red
sequence hosts.  Withing this distribution, AGN strength is correlated
to the youth of the host galaxy stellar population, such that bluer
galaxies have more luminous broad-line AGNs.  This suggests that
broad-line AGN activity and star formation are closely coeval with
little or no delay between the ignition of each.  The host galaxy
properties of broad-line AGN hosts also suggest that quenching of
galactic star formation occurs or becomes visible only after the
rapidly accreting SMBH phase: quasar winds are either unconnected to
quenching, or their effects are apparent only after the broad-line AGN
phase has ended.

\acknowledgements The authors from UCSC acknowledge support from NASA
HST grant GO 12060.10-A, Chandra grant G08-9129A, and NSF grant AST-
0808133.  Helpful discussions with Edmond Cheung and Hassen Yesuf
contributed to the development of this work.  AH acknowledges support
from the UC Santa Cruz Science Internship Program.

\LongTables
\begin{deluxetable*}{lrrrrrrrrrr}  
  \tablecolumns{11}
  \tablecaption{Raw and Corrected Measurements of Broad-Line AGN Hosts
    \tablenotemark{a} \label{tbl:catalog}}
  \tablehead{
    \colhead{SDSS ID} & 
    \colhead{RA} & 
    \colhead{Dec} & 
    \colhead{redshift} & 
    \colhead{$M_z'$\tablenotemark{b}} & 
    \colhead{Raw $u'$\tablenotemark{b}} & 
    \colhead{Cor. $u'$\tablenotemark{b}} & 
    \colhead{${\delta}u$} & 
    \colhead{$C_z$} & 
    \colhead{$(u-z)'$\tablenotemark{b}} &
    \colhead{$\log(M_*/M_\odot)$} }
  \startdata
  587732484349100048 & 147.638145 &  44.314366 & 0.0153 & -20.21 & 16.73 & 17.00 &  0.40 &  4.19 &  3.52 & 10.18 \\
  587742060556648547 & 233.968767 &  14.517718 & 0.0195 & -20.66 & 15.94 & 16.18 &  1.16 &  2.25 &  2.65 & 10.06 \\
  587735240099954710 & 129.386277 &  28.705193 & 0.0114 & -21.35 & 14.88 & 14.92 &  0.21 &  2.20 &  3.39 & 10.67 \\
  587734948589273110 & 146.372371 &   9.602890 & 0.0133 & -19.46 & 16.41 & 16.91 &  1.02 &  2.86 &  2.87 &  9.71 \\
  587739811560882185 & 214.498122 &  25.136867 & 0.0165 & -21.81 & 14.78 & 15.20 &  1.70 &  3.09 &  3.15 & 10.73 \\
  587724242842026028 &  52.555273 &  -5.543318 & 0.0131 & -20.91 & 14.91 & 14.99 &  0.35 &  2.55 &  2.81 & 10.42 \\
  587734622701093005 & 121.411097 &  26.168188 & 0.0170 & -21.52 & 15.15 & 15.21 &  0.63 &  1.90 &  2.95 & 10.62 \\
  587735695380578421 & 224.278321 &  49.669022 & 0.0134 & -20.93 & 14.57 & 14.72 &  0.99 &  2.85 &  2.26 & 10.15 \\
  587725039018311737 & 180.309815 &  -3.678079 & 0.0196 & -20.46 & 15.46 & 16.39 &  2.23 &  2.74 &  2.27 &  9.80 \\
  587738946662629527 & 129.545598 &  24.895276 & 0.0284 & -21.76 & 16.09 & 16.31 &  0.70 &  2.71 &  3.30 & 10.89 \\
  587732577773420642 & 161.215528 &   6.596847 & 0.0276 & -22.37 & 15.11 & 15.15 &  0.03 &  2.09 &  2.55 & 10.85 \\
  587738616483282976 & 154.956204 &  33.367703 & 0.0232 & -19.77 & 16.10 & 17.41 &  2.16 &  3.00 &  2.67 &  9.86 \\
  587726014003216406 & 198.274215 &   1.465540 & 0.0294 & -20.22 & 17.31 & 17.64 &  0.32 &  2.78 &  2.92 & 10.15 \\
  587733080809668634 & 171.400663 &  54.382551 & 0.0206 & -19.18 & 17.00 & 17.78 &  1.19 &  3.03 &  2.66 &  9.60 \\
  587741490361204744 & 141.827109 &  23.020102 & 0.0263 & -21.76 & 16.18 & 16.36 &  0.27 &  3.00 &  3.52 & 10.89 \\
  587732771575955522 & 144.551130 &   7.727765 & 0.0219 & -20.08 & 16.92 & 17.42 &  0.83 &  3.01 &  3.20 & 10.09 \\
  587729233053417523 & 251.839456 &  44.702715 & 0.0254 & -19.98 & 17.41 & 17.57 & -0.31 &  2.54 &  2.88 &  9.97 \\
  587733603730063381 & 245.053147 &  40.151711 & 0.0285 & -20.42 & 16.16 & 16.77 &  1.43 &  3.11 &  2.16 &  9.93 \\
  588010359086579732 & 180.740878 &   4.845853 & 0.0207 & -21.37 & 15.87 & 15.98 &  0.96 &  1.89 &  3.08 & 10.58 \\
  \enddata
  \tablenotetext{a}{The full catalog of 820 broad-line AGNs appears as
    a machine-readable table in the electronic version.  In addition
    to the columns shown here, the full catalog includes raw and
    corrected magnitudes and ${\delta}m$ measurements for all five
    $ugriz$ filters.}
  \tablenotetext{b}{All magnitudes are K-corrected to $z=0.05$ and
    given in AB units.}
\end{deluxetable*}   

\newpage
\LongTables
\begin{deluxetable*}{cc|rrr|rrr}   
  \tablecolumns{8}
  \tablecaption{Broad-Line AGN Fractions with Host Galaxy Stellar Mass
    and Color \label{tbl:agnfrac}}
  \tablehead{
    \multicolumn{2}{c}{Lower Left Bin\tablenotemark{a}} &
    \multicolumn{3}{c}{Faint Sample} &
    \multicolumn{3}{c}{Luminous Sample} \\
    \colhead{$\log(M_*/M_\odot)$} & 
    \colhead{$(u-z)'$} & 
    \colhead{$N_{\rm gal}$} & 
    \colhead{$N_{\rm AGN}$} & 
    \colhead{BL AGN \%} &
    \colhead{$N_{\rm gal}$} & 
    \colhead{$N_{\rm AGN}$} & 
    \colhead{BL AGN \%} }
  \startdata
     9.00 & 1.175 &   208 &     0 &             0   &    0 &    0 &           0   \\
     9.00 & 1.475 &   500 &     0 &             0   &    0 &    0 &           0   \\
     9.00 & 1.775 &   123 &     0 &             0   &    0 &    0 &           0   \\
     9.00 & all colors & 835 &  0 &             0   &    0 &    0 &           0   \\
     9.25 & 1.250 &   198 &     1 & $0.51 \pm 0.51$ &    0 &    0 &           0   \\
     9.25 & 1.550 &   835 &     0 &             0   &    0 &    0 &           0   \\
     9.25 & 1.850 &   979 &     1 & $0.10 \pm 0.10$ &    0 &    0 &           0   \\
     9.25 & 2.150 &   213 &     0 &             0   &    0 &    0 &           0   \\
     9.25 & 2.450 &    18 &     0 &             0   &    0 &    0 &           0   \\
     9.25 & all colors & 2246 & 2 & $0.09 \pm 0.06$ &    0 &    0 &           0   \\
     9.50 & 1.325 &   136 &     0 &             0   &    0 &    0 &           0   \\
     9.50 & 1.625 &   769 &     0 &             0   &    0 &    0 &           0   \\
     9.50 & 1.925 &  1292 &     4 & $0.31 \pm 0.16$ &    0 &    0 &           0   \\
     9.50 & 2.225 &   711 &     2 & $0.28 \pm 0.20$ &    0 &    0 &           0   \\
     9.50 & 2.525 &   330 &     2 & $0.61 \pm 0.43$ &    0 &    0 &           0   \\
     9.50 & 2.825 &   442 &     0 &             0   &    0 &    0 &           0   \\
     9.50 & 3.125 &    50 &     0 &             0   &    0 &    0 &           0   \\
     9.50 & all colors & 3731 & 8 & $0.21 \pm 0.08$ &    0 &    0 &           0   \\
     9.75 & 1.400 &    58 &     0 &             0   &   354 &     0 &             0   \\
     9.75 & 1.700 &   462 &     1 & $0.22 \pm 0.22$ &   272 &     0 &             0   \\
     9.75 & 2.000 &  1249 &     4 & $0.32 \pm 0.16$ &    22 &     0 &             0   \\
     9.75 & 2.300 &   997 &     4 & $0.40 \pm 0.20$ &     1 &     0 &             0   \\
     9.75 & 2.600 &   668 &     4 & $0.60 \pm 0.30$ &     1 &     0 &             0   \\
     9.75 & 2.900 &  1301 &     3 & $0.23 \pm 0.13$ &     0 &     0 &             0   \\
     9.75 & 3.200 &   677 &     0 &             0   &     2 &     0 &             0   \\
     9.75 & 3.500 &    31 &     0 &             0   &     2 &     0 &             0   \\
     9.75 & all colors & 5443 & 16 & $0.29 \pm 0.07$ &   654 &     0 &             0   \\
    10.00 & 1.475 &    19 &     0 &             0   &   438 &    10 & $2.28 \pm 0.73$ \\
    10.00 & 1.775 &   222 &     0 &             0   &  1929 &    18 & $0.93 \pm 0.22$ \\
    10.00 & 2.075 &   844 &     3 & $0.36 \pm 0.21$ &  1950 &    12 & $0.62 \pm 0.18$ \\
    10.00 & 2.375 &  1141 &     6 & $0.53 \pm 0.22$ &   326 &     0 &             0   \\
    10.00 & 2.675 &   821 &     8 & $0.97 \pm 0.35$ &    59 &     0 &             0   \\
    10.00 & 2.975 &  1691 &     2 & $0.12 \pm 0.08$ &     3 &     0 &             0   \\
    10.00 & 3.275 &   974 &     3 & $0.31 \pm 0.18$ &     2 &     0 &             0   \\
    10.00 & 3.575 &    50 &     0 &             0   &     1 &     0 &             0   \\
    10.00 & all colors & 5762 & 22 & $0.38 \pm 0.08$ &  4708 &    40 & $0.85 \pm 0.13$ \\
    10.25 & 1.550 &     8 &     0 &             0   &   200 &     9 & $4.50 \pm 1.53$ \\
    10.25 & 1.850 &    90 &     0 &             0   &  1510 &    31 & $2.05 \pm 0.37$ \\
    10.25 & 2.150 &   449 &     4 & $0.89 \pm 0.45$ &  4762 &    61 & $1.28 \pm 0.17$ \\
    10.25 & 2.450 &   942 &     9 & $0.96 \pm 0.32$ &  4807 &    37 & $0.77 \pm 0.13$ \\
    10.25 & 2.750 &   939 &     6 & $0.64 \pm 0.26$ &  2513 &    21 & $0.84 \pm 0.18$ \\
    10.25 & 3.050 &  1933 &     2 & $0.10 \pm 0.07$ &  2291 &     1 & $0.04 \pm 0.04$ \\
    10.25 & 3.350 &  1211 &     0 &             0   &   493 &     0 &             0   \\
    10.25 & 3.650 &    72 &     0 &             0   &     6 &     0 &             0   \\
    10.25 & all colors & 5644 & 21 & $0.37 \pm 0.08$ & 16582 &   160 & $0.96 \pm 0.08$ \\
    10.50 & 1.625 &     4 &     0 &             0   &    54 &     1 & $1.85 \pm 1.87$ \\
    10.50 & 1.925 &    16 &     0 &             0   &   511 &    21 & $4.11 \pm 0.92$ \\
    10.50 & 2.225 &   142 &     3 & $2.11 \pm 1.23$ &  2617 &    44 & $1.68 \pm 0.26$ \\
    10.50 & 2.525 &   530 &     9 & $1.70 \pm 0.57$ &  5406 &    49 & $0.91 \pm 0.13$ \\
    10.50 & 2.825 &   834 &    13 & $1.56 \pm 0.44$ &  5584 &    40 & $0.72 \pm 0.11$ \\
    10.50 & 3.125 &  1957 &     6 & $0.31 \pm 0.13$ & 11466 &    22 & $0.19 \pm 0.04$ \\
    10.50 & 3.425 &  1205 &     0 &             0   & 12747 &     1 & $0.01 \pm 0.01$ \\
    10.50 & 3.725 &    81 &     0 &             0   &  1220 &     3 & $0.25 \pm 0.14$ \\
    10.50 & all colors & 4769 & 31 & $0.65 \pm 0.12$ & 39605 &   181 & $0.46 \pm 0.03$ \\
    10.75 & 1.700 &     0 &     0 &             0   &    16 &     0 &             0   \\
    10.75 & 2.000 &     1 &     0 &             0   &   114 &     7 & $6.14 \pm 2.39$ \\
    10.75 & 2.300 &    39 &     1 & $2.56 \pm 2.60$ &   842 &    21 & $2.49 \pm 0.55$ \\
    10.75 & 2.600 &   163 &     3 & $1.84 \pm 1.07$ &  2419 &    21 & $0.87 \pm 0.19$ \\
    10.75 & 2.900 &   406 &     4 & $0.99 \pm 0.50$ &  3921 &    20 & $0.51 \pm 0.11$ \\
    10.75 & 3.200 &  1622 &     2 & $0.12 \pm 0.09$ & 10541 &    10 & $0.09 \pm 0.03$ \\
    10.75 & 3.500 &   884 &     1 & $0.11 \pm 0.11$ & 12972 &     4 & $0.03 \pm 0.02$ \\
    10.75 & 3.800 &    64 &     0 &             0   &  1615 &     2 & $0.12 \pm 0.09$ \\
    10.75 & all colors & 3179 & 11 & $0.35 \pm 0.10$ & 32440 &    85 & $0.26 \pm 0.03$ \\
    11.00 & 2.075 &     1 &     0 &             0   &    11 &     0 &             0   \\
    11.00 & 2.375 &     5 &     0 &             0   &   104 &     2 & $1.92 \pm 1.37$ \\
    11.00 & 2.675 &    16 &     0 &             0   &   536 &     9 & $1.68 \pm 0.56$ \\
    11.00 & 2.975 &   113 &     1 & $0.88 \pm 0.89$ &  1449 &     7 & $0.48 \pm 0.18$ \\
    11.00 & 3.275 &   801 &     3 & $0.37 \pm 0.22$ &  6078 &     9 & $0.15 \pm 0.05$ \\
    11.00 & 3.575 &   425 &     0 &             0   &  7530 &     1 & $0.01 \pm 0.01$ \\
    11.00 & 3.875 &    44 &     0 &             0   &   904 &     0 &             0   \\
    11.00 & all colors & 1405 & 4 & $0.28 \pm 0.14$ & 16614 &    28 & $0.17 \pm 0.03$ \\
    11.25 & 2.450 &     0 &     0 &             0   &    12 &     1 & $8.33 \pm 8.67$ \\
    11.25 & 2.750 &     2 &     0 &             0   &    33 &     0 &             0   \\
    11.25 & 3.050 &     6 &     0 &             0   &   186 &     1 & $0.54 \pm 0.54$ \\
    11.25 & 3.350 &   207 &     0 &             0   &  1769 &     1 & $0.06 \pm 0.06$ \\
    11.25 & 3.650 &   101 &     0 &             0   &  2453 &     2 & $0.08 \pm 0.06$ \\
    11.25 & 3.950 &    14 &     0 &             0   &   280 &     0 &             0   \\
    11.25 & all colors & 330 &  0 &             0   &  4737 &     5 & $0.11 \pm 0.05$ \\
    11.50 & 3.125 &     0 &     0 &             0   &     9 &     0 &             0   \\
    11.50 & 3.425 &    24 &     0 &             0   &   277 &     0 &             0   \\
    11.50 & 3.725 &     8 &     0 &             0   &   357 &     0 &             0   \\
    11.50 & 4.025 &     1 &     0 &             0   &    62 &     0 &             0   \\
    11.50 & all colors &  33 &  0 &             0   &   708 &     0 &             0   \\
  \enddata
  \tablenotetext{a}{Columns of $\log(M_*/M_\odot)$ and $(u-z)'$ give
    the values associated with the lower left corner of each
    color-mass bin, as shown in Figures \ref{fig:cmdfaint} and
    \ref{fig:cmdlumin}.  Bins are 0.25 dex wide in mass and 0.3 dex
    high in color, and bins with less than five inactive galaxies are
    not shown.  Rows with ``all colors'' give the numbers and fraction
    of broad-line AGNs in that entire mass slice.}
\end{deluxetable*}   

\end{document}